\documentclass{iopart}

\usepackage{graphicx}
\usepackage{amssymb}
\usepackage{xcolor}
\usepackage{pgf}
\usepackage[]{datetime}
\usepackage[
	backend=bibtex,
	style=numeric-comp,
	natbib=false,
	url=false,
	doi=true,
	eprint=true,
	minbibnames=5,
	maxbibnames=10,
	sorting=none
]{biblatex}

\makeatletter
\g@addto@macro\bfseries{\boldmath}
\makeatother

\newcommand{\Pabsorbed}{P_{\mathrm{optH}}}
\newcommand{\Pgas}{P_{\mathrm{gasC}}}
\newcommand{\Pvics}{P_{\mathrm{VF}}}
\newcommand{\Pvisc}{\Pvics}
\newcommand{\Pscrews}{P^\mathrm{screws}_{\mathrm{cond}}}

\newcommand{\Pdis}{\Delta P_{\mathrm{res}}}

\newcommand{\Tframe}{T_\mathrm{frame}}
\newcommand{\TTM}{T_\mathrm{TM}}

\begin{document}

\title[Thermal accommodation coefficient of helium
	on cold crystalline silicon]{Measurement of the thermal
	accommodation coefficient of helium
	on crystalline silicon at low-temperatures}

\author[Alexander Franke et. al.]{
	Alexander Franke,
	Nils Sültmann,
	Christoph Reinhardt,
	Sandy Croatto,
	Jörn Schaffran,
	Hossein Masalehdan,
	Axel Lindner,
	Roman Schnabel
}

\address{$^1$ Institut für Quantenphysik und
	Zentrum für Optische Quantentechnologien, Universität Hamburg, 22761 Hamburg, Germany}
\address{$^2$ Deutsches Elektronen-Synchrotron DESY, Notkestr. 85, 22607 Hamburg, Germany}

\ead{afranke@physnet.uni-hamburg.de}
\vspace{10pt}
\begin{indented}
	\item[]February 2024
	\item[]
	\item[]\textit{All figures and pictures by the authors under a CC BY 4.0 license}
\end{indented}

\begin{abstract}
	Proposals for next-generation gravitational wave observatories include
	cryogenically cooled 200-kg test mass mirrors
	suspended from pendulums and made of a crystalline material such as crystalline silicon.
	During operation of the observatories,
	these mirrors undergo heating due to the absorption
	of laser radiation of up to a watt.
	Low noise cooling techniques need to be developed.
	Low-pressure helium exchange gas at 5\,K might contribute to the challenging task.
	Here, we report the measurement of the helium accommodation coefficient
	$\alpha(11\,\mathrm{K}<T< 30\,\mathrm{K})$,
	which is the probability that a helium atom thermalises with a surface
	at a given temperature
	when reflected from it.
	We find $\alpha(T) > 0.7$ for temperatures $< 20$\,K,
	which increases the cooling power compared to recently used assumptions.
	The idea of free molecular flow helium gas cooling is thus supported
	and might find application in some observatory concepts.
\end{abstract}

\section{Introduction}\label{sec:introduction}
The first observation of gravitational waves (GW)~\cite{first-observation} marked the beginning
of a
new chapter in astronomy and cosmology,
as gravitational wave detection offers a new way
to study the universe alongside traditional telescope observations.
Fast-forward to today,
several improvements to the detectors~\cite{aasi2015advanced,acernese2014advanced,akutsu_kagra_2019}
have been implemented.
These sensitivity improvements contributed to many more event detections,
thereby providing valuable information about the number density of black holes~\cite{gais_inferring_2022},
the gravitational wave background~\cite{Agazie_2023},
and an independently derived value of the rate
at which the universe is expanding
(Hubble constant $H_0$)~\cite{hubble-constant-from-wave_2017}.

With next-generation interferometers~\cite{abernathy2011einstein,ET2020einstein,reitze2019cosmic,nemo,adhikari2020cryogenic,ligovoyager1,voyagerwhitepaper},
further improvements in the sensitivity are targeted.
With the detection of weaker signals,
we can potentially gain knowledge about gravity,
the structure of spacetime, supernovae,
compact binary systems long before they merge, and cosmic inflation.
To this end,
the design plans consider cooling the laser mirrors and test masses of space-time (TMs),
which are of the order of 100\,kg
and suspended as pendulums,
to cryogenic temperatures.
On the one hand, this is aimed at suppressing thermal noise,
which is currently limiting in the range from 40\,Hz to 100\,Hz~\cite{buikema2020sensitivity,acernese2020advanced}.
On the other hand, this might enable higher light powers~\cite{nemo}
thereby reducing shot noise, which is limiting above 100\,Hz.

Today the Japanese KAGRA observatory~\cite{somiya2012detector,akutsu_kagra_2019},
which began initial observations
in Feb 2020~\cite{overview-kagra}, will finally exploit mirrors cooled to about 20\,K\@.
The designs of the European
Einstein Telescope and LIGO Voyager incorporate cryo-cooling as well.
The ET pathfinder project is currently evaluating
different cryogenic configurations
for the Einstein Telescope~\cite{team2020ETpathfinder,utina_etpathfinder_2022}
aiming at either 123\,K or 18\,K\@.

The mirrors in the arm cavities are suspended on meter-scale,
thin wires,
behaving like the well-understood \emph{harmonic oscillator}:
Disturbances above the resonance frequency are significantly reduced in amplitude.
However,
this excellent mechanical isolation comes with the drawback of substantial thermal isolation,
as thin wires only conduct a very limited amount of heat.
To reduce quantum shot noise, interferometric detectors need to use high-power lasers,
which heat up mirror coatings and substrates.
This thermal noise partially limits future observatories.

In a previous article~\cite{gas-cooling},
we proposed
cooling the mirrors of a gravitational wave detector with a 5\,K-helium gas
at low pressure in
the molecular flow regime.
We showed
that the momentum transfer results in an additional mirror displacement noise
that is proportional to $f^{-2}$,
where $f$ is the GW frequency.
We concluded that gas cooling is particularly suitable for observatories
aiming to detect events in the kilohertz range,
such as the Neutron Star Extreme Matter Observatory
(NEMO)~\cite{nemo}.

The effectiveness of gas cooling depends on
the thermal accommodation coefficient $\alpha$,
which represents the fraction of the incident gas particles
that thermalize with the TM\@.
Its value depends on the type of gas, surface material, and temperature.
For helium, $\alpha=0.43$ has been measured at room temperature~\cite{trott_experimental_2011}.
At 20 K, $\alpha=0.6$ has been estimated,
based on measurements with glass, platinum, and nickel surfaces~\cite{CORRUCCINI195919}.
Here,
we measure the thermal accommodation coefficient of helium on crystalline silicon between 11\,K and
30\,K\@.

\section{Heat transfer in gases}\label{sec:theory}
The heat transfer of gases
confined in a volume is dependent on
the mean free path $\lambda$ compared to the wall-to-wall distance $d$.
The Knudsen number $K_n=\lambda d^{-1}$ defines three regimes:
viscous flow ($K_n \ll 1$),
free molecular ($K_n \gg 1$),
and transitional flow in between.

\subsection{Gas cooling in the viscous flow regime}
\label{ssec:gas-cooling-in-the-viscous-flow-regime}

At ambient pressures,
the presence of molecular collisions and momentum transfer between gas molecules
almost always dominate the behavior of gases.
This regime of gas dynamics is called \textit{viscous flow} or continuous flow.
Here, the heat transfer of gas between two identical, parallel plates of area $A$,
separated by distance $d$ is given by

\begin{equation}
	\Pvics = \frac{\kappa(T)  A \Delta T}{d} \qquad \mathrm{if\, \lambda \ll d},
	\label{eq:viscous-cooling}
\end{equation}

with the temperature-dependent thermal conductivity of the gas $\kappa$,
and temperature difference of the plates $\Delta T$.

For $\lambda \approx d$,
wall-molecule-interactions start to influence heat transfer.
For $\lambda \gg d$, gas molecules rarely collide with each other,
leading to more frequent thermal energy transfers with the walls.

\subsection{Gas cooling in the free molecular flow regime}
\label{subsec:gas-cooling-in-the-free-molecular-flow-regime}
The heat transfer in the free molecular flow is additionally influenced by the surface material.
The energy accommodation coefficient $\alpha \in [0,1]$ accumulates these properties in a
probability:
Complete thermal equilibration at every molecule-surface interaction translates to $\alpha = 1$,
no thermal exchange at any interaction translates to $\alpha=0$~\cite{CORRUCCINI195919}.
This coefficient is dependent on the gas type and surface material,
and varies with temperature.

With the assumption of the colder plate satisfying $\alpha = 1$,
the gas cooling power
in the free molecular flow regime
is given by~\cite{gombosi1994gaskinetic}:

\begin{equation}
	P_{\mathrm{fMF}}=\alpha(T)\sqrt{\frac{8k_{\mathrm{B}}}{\pi m_{\mathrm{Gas}}T_{\mathrm{cold}}}} p A \Delta T
	\label{eq:fmf_cooling_power}
\end{equation}

with energy accommodation coefficient of the hot surface $\alpha$,
mass of a gas molecule $m_\mathrm{Gas}$,
gas pressure at the colder surface $p$, and area of a single plate $A$.
For copper surfaces and temperatures significantly below 10\,K,
the accommodation coefficient is close to unity~\cite{CORRUCCINI195919}.

\section{Setup for measuring $\alpha_\mathrm{He,cSi}(11\,\mathrm{K}<T<30\,\mathrm{K})$}
\label{sec:setup}

\begin{figure}
	\centering
	\includegraphics[width=.99\linewidth]{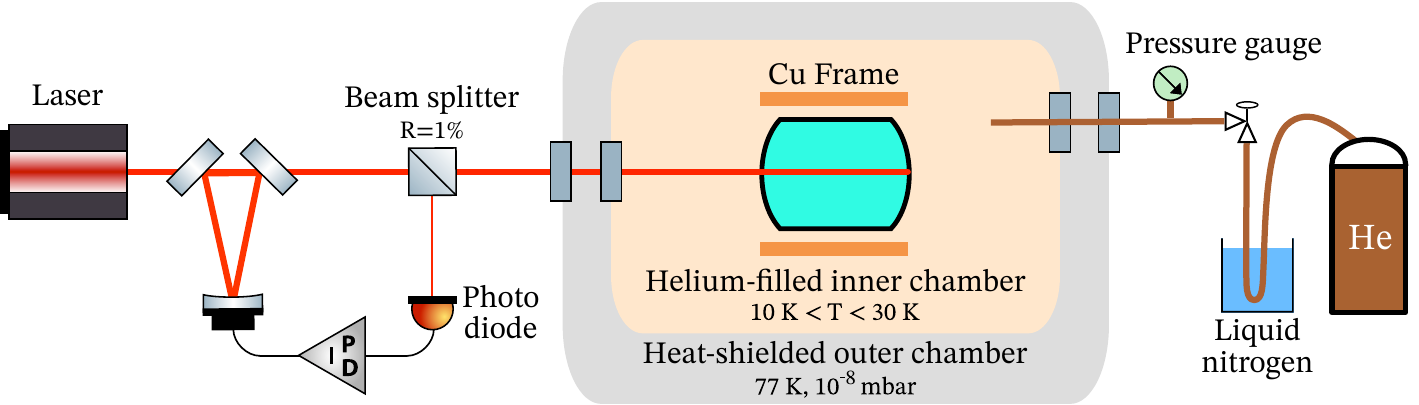}
	\caption{Simplified experimental setup.
		In order to measure $\alpha_{\mathrm{He,cSi}}(11\,\mathrm{K}<T<30\,\mathrm{K})$, we place
		a cylindrical Si crystal (cyan) inside a helium-tight chamber
		(light orange), which is thermally connected to a cryogenically cooled plate.
		A power-adjustable laser (red) is guided through a pre-mode cleaner and a fiber for controlled heating of the test mass.
		We stabilize the light power, by splitting off 1\% of power
		and directing it onto a photodiode.
		The output signal from the photodiode is used
		to control the light power
		allowed to pass the pre-mode-cleaner, which achieves reasonable power stability
		by using a variable offset DC-Lock.
		Optical power seen by the silicon TM fluctuates $<0.6\,\%$ during measurements.
		See section~\ref{ssec:calibration-of-the-absorbed-optical-power}
		for a description of the calibration of the absorbed power
		inside the test mass.
		Helium from a pressurized gas bottle (brown)
		with purity of 99.999\,\% is guided through liquid nitrogen
		to remove contaminations.
		A variable leak rate valve controls the gas pressure measured by the gauge (green).
		A capillary connects inner chamber and pressure gauge.  
	}
	\label{fig:optical-setup}
\end{figure}

A cylindrical, crystalline silicon test mass
was placed inside an indium-sealed,
vacuum chamber with a wall temperature of 10\,K or less and adjustable helium pressure.
A laser enabled us to deposit a known, tunable amount of heat energy in the test mass.
To reach cryogenic temperatures,
the aforementioned \emph{inner chamber} is mounted to a cold plate inside a bigger \emph{outer chamber},
which could be pumped to pressures below $10^{-8}$\,mbar.
A thermal shield surrounded the inner chamber to reduce radiative heating.
This setup (shown in more detail in Figure~\ref{fig:optical-setup})
enabled us
to reach silicon surface temperatures of 11 to 30\,K,
which covers most of the suggested cryogenic detector designs mentioned in the introduction.

We used a crystalline silicon cylinder made with the Czochralski
technique~\cite{czochralski1918neues}
that was cut to have a diameter of ($2.4863\pm0.0024$)\,cm and a length of ($9.972\pm0.004$)\,cm.
The test mass was held inside a polished copper frame almost fully enclosing it.
The frame was fabricated to be equidistant to all silicone surfaces.
This controlled distance of 2\,mm is necessary to ensure compatibility
with the free molecular flow regime requirement.
Differences in thermal expansion of copper and silicon are insignificant~\cite{copper1,silicon-expansion}.
Six polytetrafluoroethylene (PTFE) screws with pointy tips held the test mass in the center
of the frame.
PTFE is known for its low thermal conductivity~\cite{teflonconductivity}.
Thermal sensors were mounted to the test mass and the frame
to measure their respective temperatures.
Without any additional heat load, the test mass could be cooled to 7\,K\@ with similar frame temperatures.

Both chambers had optical ports.
These allowed pointing a laser beam at wavelength 1064\,nm onto the test mass.
A capillary thermally anchored over a length of around 80\,cm
to the heat shield connects the inner chamber to the gas handling at room temperature.
The in-going laser power was held stable by a control loop with long-term stability.
Additionally, this control loop allowed us
to automatically heat the test mass to a desired temperature.
Once the system reached equilibrium, data was collected for up to 30 minutes,
where the following equation connects optical heating power $\Pabsorbed$ and gas cooling power $\Pgas$:
\begin{equation}
	\Pabsorbed = \Pgas + \Pdis
	\label{eq:fundamental_power_relation}
\end{equation}
where $\Pdis$ contains any remaining heat flow channels.
$\Pabsorbed$ is proportional to the input laser power.
The light-power calibration is described in
section~\ref{ssec:calibration-of-the-absorbed-optical-power}.
The value of $\Pgas$ is dependent on the gas pressure,
and in the case of molecular flow:
the accommodation coefficient.
$\Pdis$ is discussed in section~\ref{ssec:discommoding-cooling}.
A front view of the system in the inner chamber is shown in Figure~\ref{fig:holder-tm-setup}.

\section{Measurement of the thermal accommodation coefficient}\label{sec:results}

\subsection{Pressure measurement}\label{ssec:pressure-sensing}
The experiment operated in three gas pressure regimes:
no-gas ($K_n \gg 10$),
gas in molecular flow ($8 < K_n < 12$) and
gas in viscous flow ($K_n < 0.01$).
During individual measurements,
helium inlet valve was closed;
hence the pressure was static.
The first and last regimes are needed for calibration:
Measurement without gas allows
for the determination of any unwanted heat flux
(see section~\ref{ssec:discommoding-cooling}).
Gas-cooling power in viscous flow is known
(see~equation~\ref{eq:viscous-cooling}) and
enables a precise calibration of the applied heating power
(see section~\ref{ssec:calibration-of-the-absorbed-optical-power}).
With that,
cooling power in molecular flow is proportional to one last unknown:
the accommodation coefficient.

At the state of planning,
no commercial pressure sensor for cryogenic temperatures was available to us.
As a workaround,
we connected the inner chamber via a capillary
to a gas-type independent pressure gauge outside the cryostat.
This capillary introduced a small heat bridge
and therefore limited the minimum achievable temperature.
Due to the \emph{thermomolecular pressure difference effect},
a correction to the pressure readout needed to be applied.
Given that the gas in the capillary was in molecular flow,
the pressure readout in the warm region $p_w$ at temperature $T_w$ could be measured
and translated to the pressure inside the cryostat $p_c$ at temperature $T_c$ with:
\begin{equation}
    \frac{p_c}{p_w} = \sqrt{\frac{T_c}{T_w}}\label{eq:pressure-temp-correction}
\end{equation}
At temperatures of 8\,K, the pressure inside the inner chamber was approximately six times lower
than at the pressure gauge.
The required room temperature measurements added another small uncertainty of 3--4\,\% to the
determined accommodation coefficients.

For a gas-type independent readout,
we used a commercially available capacitive pressure transmitter.
By design, these pressure gauges have the smallest uncertainty
at the upper end of their sensitivity range.
Consequently,
we operated the experiment between $K_n = 8$ and $K_n = 12$
during molecular flow measurements.
In testing,
we found that zeroing of the sensor every 48 hours
is necessary as the readout value drifted.
Zeroing had to be executed at low pressures, in our case at $p < 10^{-6}\,$mbar.
Hence, the gas filled part of the experiment outside the cryostat gets pumped
and flushed with fresh helium every two days.
This had the added benefit that we ensured high-level purity of helium at all times.

\subsection{Temperature measurement}
\label{subsec:calibration-of-temperature-sensors}
\begin{figure}
    \centering
    \includegraphics[width=.42\linewidth]{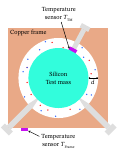}
    \caption{
        Cross section of the silicon cylinder (cyan) and its surrounding copper frame (orange).
        The silicon cylinder is held in place by six PTFE screws (light gray) with a distance to
        the frame of d\,=\,2\,mm.
        PTFE was chosen for its low thermal conductivity.
        For a similar reason, the tips are pointy.
        Thin film resistance sensors (purple) allow precise temperature readout.
        A seventh screw pushes one of the sensors onto the test mass.
        Alongside thermal grease,
        this allows for great thermal contact
        to ensure the sensor shows the test mass temperature
        as precisely as possible.
        Helium atoms (small red and blue dots)
        with different temperatures ($\TTM$ and $\Tframe$) at a pressure $p_{c}$
        exchange heat between frame and test mass.
    }
    \label{fig:holder-tm-setup}
\end{figure}

Temperatures of the test mass ($\TTM$) and surrounding frame ($\Tframe$)
were read out with \emph{negative temperature coefficient (NTC) resistor sensors}.
The manufacturer claims readout error of 4\,mK between 4\,K to 10\,K and 8\,mK between 10\,K to 20\,K\@.
A sensor (model Cernox 1070 HT sensor in CU package)
was attached to the copper frame.
Thermal grease and a hand-tight brass screw ensured good thermal contact.
The cylindrical test mass was thermally well insulated from the cold copper.
Its sensor (model Cernox 1070 HT sensor in SD package)
may not be fixed to a flat surface,
as laser light would heat it up if placed on one end of the cylinder.
We chose a sensor
assembled in a 3\,mm $\times$ 2\,mm rectangular package with about 1\,mm height.
A PTFE screw with bad thermal conductance pressed
the flat surface of the sensor onto the curved test mass (see Figure~\ref{fig:holder-tm-setup}).

The temperature sensor controller unit applied
temperature-dependent excitation current to each sensor
to read out their electrical resistance.
The readout showed only the voltage drop over the sensor,
as we used a four-wire scheme making wire resistance irrelevant.
The controller converted voltage readout
to temperature with a manufacturer-provided calibration.
A small electrical self-heating at each sensor was technically unavoidable.
This heating was notable at the test mass sensor,
as it was highly insulated from the cold plate of the cryostat.
The electrical self-heating power was temperature-dependent,
as with changing sensor resistance,
the electrical current is also changing.

In the no-gas (molecular flow) [viscous flow] regime,
we observed stable temperature readout differences between test mass and frame sensor of 757\,mK
(36\,mK) [31\,mK].
These differences were measured
when all optical accesses to the test mass were blocked
and no heat load other than sensor heating was present at the test mass.
This showed a strong dependence of the temperature differences on gas pressure.
Gas was exchanging heat between the test mass sensor
and both the frame and test mass.
Several numerical and experimental tests after the experiment run revealed
that the readout of the test mass sensor has a high dependence
on the specific mounting procedure and gas pressure.
We suspect a heat accumulation inside the sensor,
which was most notable in the no gas
configuration, as the helium greatly increases the thermal contact
between resistor and test mass surface.
To compensate, we subtracted test mass temperature readings
by above-mentioned values in dependence of the gas regime.
At the same time, we continued with a systematic uncertainty equally to the subtracted amount.
This fact had an influence on our no-gas calibration measurements determining $\Pdis$,
and the large uncertainty on temperature measurements
during the calibration
can be seen Figure~\ref{fig:supplementary-plots} right.
Error propagation dictated higher uncertainties in the accommodation coefficient for
experiment configurations at temperatures below 14\,K\@.

\subsection{Calibration of the absorbed optical power}
\label{ssec:calibration-of-the-absorbed-optical-power}
A laser of wavelength 1064\,nm deposited an adjustable
and stable amount of thermal energy inside the test mass.

The end surfaces of the cylinder were polished and coated.
For this experiment,
it is irrelevant if the light is absorbed in the coatings or the bulk.
COMSOL Multiphysics simulations have shown
that the temperature inside the test mass is always nearly uniform,
due to the high thermal conduction of silicon.

A beam splitter split off 1\,\% of laser light just before the vacuum chamber
which we guided onto a photodiode.
The photovoltage provided feedback to a pre-mode-cleaner
to compensate for thermal drifts in the laser system
and therefore to ensure a stable heating power.

\begin{figure}
    \centering
    \includegraphics[width=.49\textwidth]{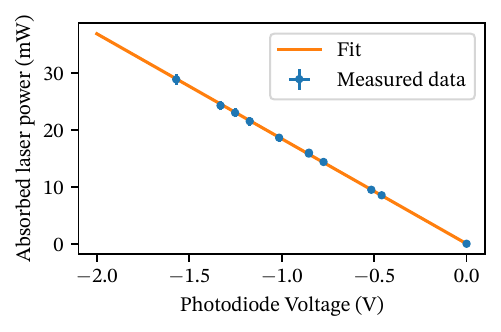}
    \includegraphics[width=.49\textwidth]{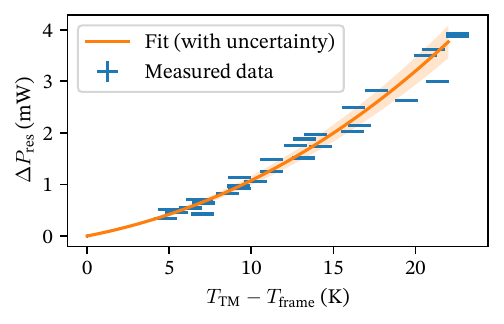}
    \caption{\textbf{Left: }
    A small, fixed portion of the laser light heating the test mass, is directed on a photodiode.
    The linear correlation between the voltage readout of this diode
    and absorbed laser power in the test mass is shown here.
    Data taken with gas in the viscous flow regime, where gas cooling is dominating
    and known (eq.~\ref{eq:viscous-cooling}).
        \textbf{Right:} Residual cooling power of every heat flux channel
        apart from gas cooling $\Pdis$.
        This is dominated by the heat conduction through the screws
        holding the test mass in place.
        Measured at pressures well below $10^{-6}$\,mbar,
        where no cooling by gas contributes significantly
        to the heat flux of the test mass.
        The high temperature uncertainty is explained
        in section~\ref{subsec:calibration-of-temperature-sensors}.
    }
    \label{fig:supplementary-plots}
\end{figure}

We determined the exact amount of optical heating
applied to the test mass $\Pabsorbed$.
Measuring incoming, reflected,
and transmitted optical power proved to be nearly impossible,
especially as the test mass does not form a cavity and optical scattering
far off the laser axis occurred.
The test mass position could not be adjusted in its position or angle
once the vacuum chamber was closed
(see Figure~\ref{fig:optical-setup}).
However,
we correlated the photovoltage (used to stabilize laser power)
to the actual light power absorbed in the test mass.
The cooling performance of helium is well known in the regime of viscous flow
(see equation~\ref{eq:viscous-cooling}).
If we assume viscous cooling $\Pvisc$ dominates
and the test mass temperature is constant,
we can directly calibrate photovoltage with $\Pgas = \Pvics = \Pabsorbed$.
The results of this calibration are shown in Figure~\ref{fig:supplementary-plots} (left).

\subsection{Unwanted heat flux}
\label{ssec:discommoding-cooling}
We could not mount our test mass in such a way,
that it was thermally fully decoupled from its environment,
and cooling channels apart from gas-cooling remained.
Our test mass was held in place by screws with low thermal conductivity,
but nevertheless, some Milliwatts were cooled by conduction through these.
An upper limit for
heating power by thermal conduction through $N_s$ screws of
diameter $d_s$,
temperature dependent thermal conductivity of teflon $\kappa_s(T)$~\cite{teflonconductivity},
and length $l$ is given by~\cite{ekin_experimental_2006}:
\begin{equation}
    \Pscrews = \frac{N_s \kappa_s(T)}{l}\frac{\pi d_s^2}{4} \Delta T.
    \label{eq:heat-transfer-conductive}
\end{equation}
Contact resistance and pointy tips reduced the actual conductivity.

Other thermal channels included radiation cooling to the copper frame,
sensor self-heating,
and radiation heating from the cryostat's heat shield (77\,K).
For our measurements with helium in the free molecular flow regime,
the sum of all discommoding heat flux channels $\Pdis$ needed to be determined,
to not miscalculate gas-cooling efficiency and hence the accommodation coefficient.

With precise knowledge of the absorbed laser power $\Pabsorbed$
(see section~\ref{ssec:calibration-of-the-absorbed-optical-power})
we determined any discommoding
cooling with the helium chamber at pressure of way below $10^{-6}$\,mbar.
In this pressure regime, $\Pgas = 0$ held true,
and at constant test mass temperature, we had $\Pdis = \Pabsorbed$.
The results are shown in Figure~\ref{fig:supplementary-plots} (right).

\subsection{Extracting the accommodation coefficient}\label{ssec:extracting-the-accommodation-coefficient}

Accommodation coefficients were measured
with gas pressures in the molecular flow regime.
From~equation~\ref{eq:fundamental_power_relation} we have
\begin{equation}
    \Pabsorbed = P_\mathrm{fMF} + \Pdis.
\end{equation}
with the free molecular flow cooling power $P_\mathrm{fMF}$.
Rearranging~equation~\ref{eq:fmf_cooling_power} yields
\begin{equation}
    \alpha_\mathrm{Si} =
    \frac{\Pabsorbed-\Pdis}{ p_c A \Delta T}\cdot\sqrt{\frac{\pi
    m_{\mathrm{helium}}\Tframe}{8k_{\mathrm{B}}}}\label{eq:accommodation-coeffcient}
\end{equation}
with $\Delta T = \TTM-\Tframe$, the gas pressure around the test mass $p_c$,
surface area of the test mass $A$,
molar mass of Helium $m_\mathrm{helium}$,
and Boltzmann constant $k_B$.

\begin{figure}
    \centering
    \includegraphics[width=\textwidth]{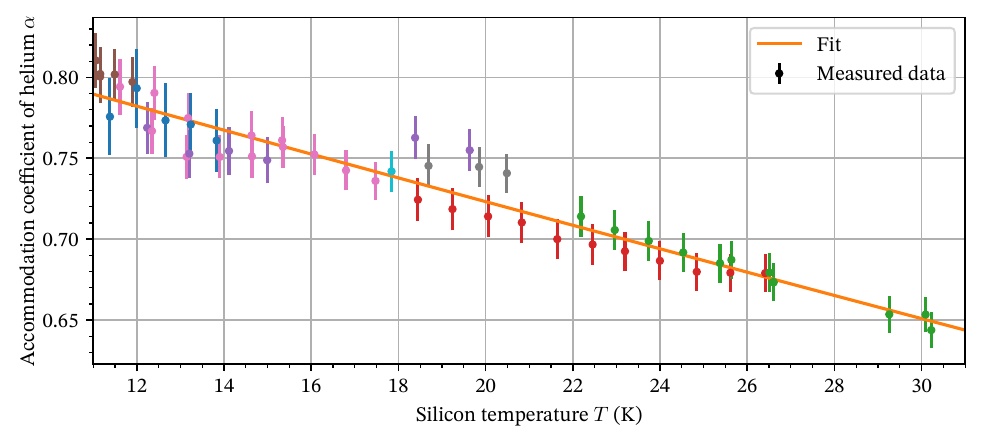}
    \caption{Measured accommodation coefficients of helium on a silicon surface at temperature $T$.
    Result of orthogonal distance regression in orange with
        $\alpha (T)=a + e^{b(T+c)} / (1 + e^{b(T+c)})$.
        \input{content/acc-fit}.
        The color coding of the measured data points represents consecutive measurement runs
        only changing test mass temperature by adjusting laser power.
        In between runs, the pressure gauge was zeroed,
        and all gas in the room temperature division of the experiment was exchanged with fresh
        helium
        to ensure that no air contamination was manipulating the pressure readout.
        Additionally, we adjusted helium pressure
        to be compatible with the free molecular flow (Knudsen number $> 8$) but as high as
        possible to reduce readout uncertainty.
        \label{fig:acc}
    }
\end{figure}

Evaluating the above equation for different laser input powers
yields the accommodation for different silicon temperatures.
Measured data points alongside a fit are shown in~Figure~\ref{fig:acc}.

\begin{figure}
    \centering
    \includegraphics[width=\textwidth]{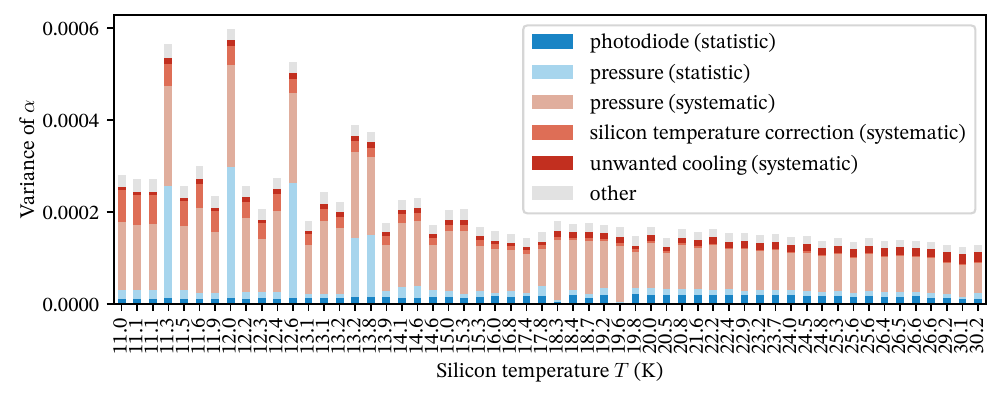}
    \caption{Uncertainites corresponding to data in Figure~\ref{fig:acc}.
    Plotted is the squared standard deviation.
    ``Other'' is the sum of variances from the following sources:
    test mass temperature sensor (statistic and systematic),
        frame temperature sensor (statistic and systematic),
        laboratory temperature,
        test mass diameter and thickness,
        and photodiode calibration (see section~\ref{ssec:calibration-of-the-absorbed-optical-power}).}
    \label{fig:uncertainties}
\end{figure}

Uncertainties in our measurements of the accommodation coefficient are dominated by a
systematic error in the pressure readout, which is given by the technical limitations of the
gauge.
A detailed breakdown of all uncertainties is shown in Figure~\ref{fig:uncertainties}.

\section{Conclusion}\label{sec:conclusion}

We have designed and executed an experiment for determination of the helium accommodation
coefficient
at cryogenic temperatures.
We covered
the silicon temperature of 18\,K, where silicon
has a zero thermal expansion coefficient
resulting in negligible thermo-elastic thermal noise in GW observatories~\cite{10.1117/12.459019}.
The accommodation coefficient
for helium on silicon cryogenic test masses has values greater than 0.7 below 20\,K\@.
Notably, for 18\,K we found $\alpha = 0.74 \pm 0.03$.

Our conceptual setup
of using a heat exchange gas to cool suspended test mass mirror in future GW observatories~\cite{gas-cooling}, assumed an accommodation coefficient of 0.6.
This work shows that for silicon temperatures of 18\,K gas cooling is approximately 23\,\% more efficient than previously discussed.
In the context of optimizing ground-based interferometric detectors like the Einstein Telescope,
this study highlights the potential benefits
of incorporating gas cooling as an additional tool in the design process.
By reducing building costs or increasing sensitivity and robustness,
such a system could contribute
to advancing our ability
to detect gravitational waves.

The potential of gas cooling in third-generation gravitational wave detectors is increased,
and
the consideration of implementing this technique should be continued,
especially for mirror masses greater than 200\,kg or detectors
targeting signal frequencies above a few hundreds of hertz.

\section*{Acknowledgements}
This work was supported and partly financed (AF, NS)
by the DFG under Germany’s Excellence Strategy EXC 2121 ``Quantum Universe'' –
390833306.
Figures created with matplotlib~\cite{Hunter:2007} and the python package uncertainties~\cite{uncertainties}.
\printbibliography

@article{first-observation,
  title = {Observation of Gravitational Waves from a Binary Black Hole Merger},
  author = {Abbott, B. P. and Abbott, R. and Abbott, T. D. and Abernathy, M. R. and Acernese, F. and Ackley, K. and Adams, C. and Adams, T. and Addesso, P. and Adhikari, R. X. and Adya, V. B. and Affeldt, C. and Agathos, M. and Agatsuma, K. and Aggarwal, N. and Aguiar, O. D. and Aiello, L. and Ain, A. and Ajith, P. and Allen, B. and Zuraw, S. E. and Zweizig, J.},
  collaboration = {LIGO Scientific Collaboration and Virgo Collaboration},
  journal = {Phys. Rev. Lett.},
  volume = {116},
  issue = {6},
  pages = {061102},
  numpages = {16},
  year = {2016},
  month = {Feb},
  publisher = {American Physical Society},
  doi = {10.1103/PhysRevLett.116.061102},
  url = {https://link.aps.org/doi/10.1103/PhysRevLett.116.061102}
}

@article{CORRUCCINI195919,
title = {Gaseous heat conduction at low pressures and temperatures},
journal = {Vacuum},
volume = {7-8},
pages = {19-29},
year = {1959},
issn = {0042-207X},
doi = {10.1016/0042-207X(59)90766-3},
url = {https://www.sciencedirect.com/science/article/pii/0042207X59907663},
author = {R.J. Corruccini}
}

@article{gais_inferring_2022,
    title = {Inferring the Intermediate-mass Black Hole Number Density from Gravitational-wave Lensing Statistics},
    volume = {932},
    issn = {2041-8205, 2041-8213},
    url = {https://iopscience.iop.org/article/10.3847/2041-8213/ac7052},
    doi = {10.3847/2041-8213/ac7052},
    pages = {L4},
    number = {1},
    journaltitle = {The Astrophysical Journal Letters},
    shortjournal = {{ApJL}},
    author = {Gais, J. and Ng, K. K. Y. and Seo, E. and Wong, K. W. K. and Li, T. G.
 F.},
    urldate = {2023-11-10},
    date = {2022-06-01},
}

@article{gas-cooling,
author = {Reinhardt, Christoph and Fran\-ke, A\-lex\-ander and Schaff\-ran, Jörn and Schna\-bel, Ro\-man and Lind\-ner, Axel},
year = {2021},
month = {09},
pages = {},
title = {Gas cooling of test masses for fu\-ture gra\-vi\-ta\-tional-wave ob\-ser\-va\-tories},
volume = {38},
journal = {Classi\-cal and Quan\-tum Gra\-vity},
doi = {10.1088/1361-6382/ac18bc}
}

@article{czochralski1918neues,
  title={Ein neues Verfahren zur Messung der Kristallisationsgeschwindigkeit der Metalle},
  author={Czochralski, Jan},
  journal={Zeitschrift f{\"u}r physikalische Chemie},
  volume={92},
  number={1},
  pages={219--221},
  year={1918},
  publisher={Oldenbourg Wissenschaftsverlag}
}

@inproceedings{10.1117/12.459019,
author = {Sheila Rowan and Robert L. Byer and Martin M. Fejer and Roger K. Route and Gianpietro Cagnoli and David R.M. Crooks and James Hough and Peter H. Sneddon and Walter Winkler},
title = {{Test mass materials for a new generation of gravitational wave detectors}},
volume = {4856},
booktitle = {Gravitational-Wave Detection},
editor = {Peter Saulson and Adrian M. Cruise},
organization = {International Society for Optics and Photonics},
publisher = {SPIE},
pages = {292 -- 297},
year = {2003},
doi = {10.1117/12.459019},
URL = {https://doi.org/10.1117/12.459019}
}

@book{ekin_experimental_2006,
	edition = {1},
	title = {Experimental {Techniques} for {Low}-{Temperature} {Measurements}: {Cryostat} {Design}, {Material} {Properties} and {Superconductor} {Critical}-{Current} {Testing}},
	isbn = {978-0-19-857054-7 978-0-19-171771-0},
	shorttitle = {Experimental {Techniques} for {Low}-{Temperature} {Measurements}},
	url = {https://academic.oup.com/book/32506},
	language = {en},
	urldate = {2023-10-23},
	publisher = {Oxford University Press},
	author = {Ekin, Jack},
	month = oct,
	year = {2006},
	doi = {10.1093/acprof:oso/9780198570547.001.0001},
}

@article{utina_etpathfinder_2022,
	title = {{ETpathfinder}: a cryogenic testbed for interferometric gravitational-wave detectors},
	volume = {39},
	url = {https://dx.doi.org/10.1088/1361-6382/ac8fdb},
	doi = {10.1088/1361-6382/ac8fdb},
	pages = {215008},
	number = {21},
	journaltitle = {Classical and Quantum Gravity},
	author = {Utina, A. and Amato, A. and Arends, J. and Arina, C. and Baar, M. de and Baars, M. and Baer, P. and Bakel, N. van and Beaumont, W. and Bertolini, A. and Beuzekom, M. van and Biersteker, S. and Binetti, A. and Brake, H. J. M. ter and Bruno, G. and Bryant, J. and Bulten, H. J. and Busch, L. and Cebeci, P. and Collette, C. and Cooper, S. and Cornelissen, R. and Cuijpers, P. and Dael, M. van and Danilishin, S. and Diksha, D. and Doesburg, S. van and Doets, M. and Elsinga, R. and Erends, V. and Erps, J. van and Freise, A. and Frenaij, H. and Garcia, R. and Giesberts, M. and Grohmann, S. and Haevermaet, H. Van and Heijnen, S. and Heijningen, J. V. van and Hennes, E. and Hennig, J.-S. and Hennig, M. and Hertog, T. and Hild, S. and Hoffmann, H.-D. and Hoft, G. and Hopman, M. and Hoyland, D. and Iandolo, G. A. and Ietswaard, C. and Jamshidi, R. and Jansweijer, P. and Jones, A. and Jones, P. and Knust, N. and Koekoek, G. and Koroveshi, X. and Kortekaas, T. and Koushik, A. N. and Kraan, M. and Kraats, M. van de and Kranzhoff, S. L. and Kuijer, P. and Kukkadapu, K. A. and Lam, K. and Letendre, N. and Li, P. and Limburg, R. and Linde, F. and Locquet, J.-P. and Loosen, P. and Lueck, H. and Martínez, M. and Masserot, A. and Meylahn, F. and Molenaar, M. and Mow-Lowry, C. and Mundet, J. and Munneke, B. and Nieuwland, L. van and Pacaud, E. and Pascucci, D. and Petit, S. and Ranst, Z. Van and Raskin, G. and Recaman, P. M. and Remortel, N. van and Rolland, L. and Roo, L. de and Roose, E. and Rosier, J. C. and Ryckbosch, D. and Schouteden, K. and Sevrin, A. and Sider, A. and Singha, A. and Spagnuolo, V. and Stahl, A. and Steinlechner, J. and Steinlechner, S. and Swinkels, B. and Szilasi, N. and Tacca, M. and Thienpont, H. and Vecchio, A. and Verkooijen, H. and Vermeer, C. H. and Vervaeke, M. and Visser, G. and Walet, R. and Werneke, P. and Westhofen, C. and Willke, B. and Xhahi, A. and Zhang, T.},
	year = {2022},
	journal = {{IOP} Publishing},
}

@article{Agazie_2023,
doi = {10.3847/2041-8213/acdac6},
url = {https://dx.doi.org/10.3847/2041-8213/acdac6},
year = {2023},
month = {jun},
publisher = {The American Astronomical Society},
volume = {951},
number = {1},
pages = {L8},
author = {Agazie, G. and Anumarlapudi, A. and Archibald, A. M. and Arzoumanian, Z. and
Baker, P. T. and Bence Bécsy and Laura Blecha and Adam Brazier and Paul R. Brook and Sarah
Burke-Spolaor and Rand Burnette and Robin Case and Maria Charisi and Shami Chatterjee and Katerina Chatziioannou and Belinda D. Cheeseboro and Siyuan Chen and Tyler Cohen and James M. Cordes and Neil J. Cornish and Fronefield Crawford and H. Thankful Cromartie and Kathryn Crowter and Curt J. Cutler and Megan E. DeCesar and Dallas DeGan and Paul B. Demorest and Heling Deng and Timothy Dolch and Brendan Drachler and Justin A. Ellis and Elizabeth C. Ferrara and William Fiore and Emmanuel Fonseca and Gabriel E. Freedman and Nate Garver-Daniels and Peter A. Gentile and Kyle A. Gersbach and Joseph Glaser and Deborah C. Good and Kayhan Gültekin and Jeffrey S. Hazboun and Sophie Hourihane and Kristina Islo and Ross J. Jennings and Aaron D. Johnson and Megan L. Jones and Andrew R. Kaiser and David L. Kaplan and Luke Zoltan Kelley and Matthew Kerr and Joey S. Key and Tonia C. Klein and Nima Laal and Michael T. Lam and William G. Lamb and T. Joseph W. Lazio and Natalia Lewandowska and Tyson B. Littenberg and Tingting Liu and Andrea Lommen and Duncan R. Lorimer and Jing Luo and Ryan S. Lynch and Chung-Pei Ma and Dustin R. Madison and Margaret A. Mattson and Alexander McEwen and James W. McKee and Maura A. McLaughlin and Natasha McMann and Bradley W. Meyers and Patrick M. Meyers and Chiara M. F. Mingarelli and Andrea Mitridate and Priyamvada Natarajan and Cherry Ng and David J. Nice and Stella Koch Ocker and Ken D. Olum and Timothy T. Pennucci and Benetge B. P. Perera and Polina Petrov and Nihan S. Pol and Henri A. Radovan and Scott M. Ransom and Paul S. Ray and Joseph D. Romano and Shashwat C. Sardesai and Ann Schmiedekamp and Carl Schmiedekamp and Kai Schmitz and Levi Schult and Brent J. Shapiro-Albert and Xavier Siemens and Joseph Simon and Magdalena S. Siwek and Ingrid H. Stairs and Daniel R. Stinebring and Kevin Stovall and Jerry P. Sun and Abhimanyu Susobhanan and Joseph K. Swiggum and Jacob Taylor and Stephen R. Taylor and Jacob E. Turner and Caner Unal and Michele Vallisneri and Rutger van Haasteren and Sarah J. Vigeland and Haley M. Wahl and Qiaohong Wang and Caitlin A. Witt and Olivia Young and The NANOGrav Collaboration},
title = {The NANOGrav 15 yr Data Set: Evidence for a Gravitational-wave Background},
journal = {The Astrophysical Journal Letters},
}

@article{hubble-constant-from-wave_2017,
	title = {A gravitational-wave standard siren measurement of the {Hubble} constant},
	volume = {551},
	copyright = {© 2017 Macmillan Publishers Limited, part of Springer Nature. All rights reserved.},
	issn = {0028-0836},
	url = {https://www.nature.com/nature/journal/v551/n7678/full/nature24471.html},
	doi = {10.1038/nature24471},
	language = {en},
	number = {7678},
	journal = {Nature},
	author = {{The LIGO Scientific Collaboration and The Virgo Collaboration} and {The 1M2H Collaboration} and {The Dark Energy Camera GW-EM Collaboration and the DES Collaboration} and {The DLT40 Collaboration} and {The Las Cumbres Observatory Collaboration} and {The VINROUGE Collaboration} and {The MASTER Collaboration}},
	month = nov,
	year = {2017},
	pages = {85--88},
}

@article{overview-kagra,
    author = {Akutsu, T. and Ando, M. and Arai, K. and Arai, Y. and Araki, S. and Araya, A. and
    Aritomi, N. and Aso, Y. and Bae, S. and Bae, Y. and Baiotti, L. and Bajpai, R. and Barton, M A
 and Cannon, K and Capocasa, E and Chan, M and Chen, C and Chen, K and Chen, Y and Chu, H and Chu, Y -K and Eguchi, S and Enomoto, Y and Flaminio, R and Fujii, Y and Fukunaga, M and Fukushima, M and Ge, G and Hagiwara, A and Haino, S and Hasegawa, K and Hayakawa, H and Hayama, K and Himemoto, Y and Hiranuma, Y and Hirata, N and Hirose, E and Hong, Z and Hsieh, B H and Huang, C -Z and Huang, P and Huang, Y and Ikenoue, B and Imam, S and Inayoshi, K and Inoue, Y and Ioka, K and Itoh, Y and Izumi, K and Jung, K and Jung, P and Kajita, T and Kamiizumi, M and Kanda, N and Kang, G and Kawaguchi, K and Kawai, N and Kawasaki, T and Kim, C and Kim, J C and Kim, W S and Kim, Y -M and Kimura, N and Kita, N and Kitazawa, H and Kojima, Y and Kokeyama, K and Komori, K and Kong, A K H and Kotake, K and Kozakai, C and Kozu, R and Kumar, R and Kume, J and Kuo, C and Kuo, H -S and Kuroyanagi, S and Kusayanagi, K and Kwak, K and Lee, H K and Lee, H W and Lee, R and Leonardi, M and Lin, L C -C and Lin, C -Y and Lin, F -L and Liu, G C and Luo, L -W and Marchio, M and Michimura, Y and Mio, N and Miyakawa, O and Miyamoto, A and Miyazaki, Y and Miyo, K and Miyoki, S and Morisaki, S and Moriwaki, Y and Nagano, K and Nagano, S and Nakamura, K and Nakano, H and Nakano, M and Nakashima, R and Narikawa, T and Negishi, R and Ni, W -T and Nishizawa, A and Obuchi, Y and Ogaki, W and Oh, J J and Oh, S H and Ohashi, M and Ohishi, N and Ohkawa, M and Okutomi, K and Oohara, K and Ooi, C P and Oshino, S and Pan, K and Pang, H and Park, J and Arellano, F E Peña and Pinto, I and Sago, N and Saito, S and Saito, Y and Sakai, K and Sakai, Y and Sakuno, Y and Sato, S and Sato, T and Sawada, T and Sekiguchi, T and Sekiguchi, Y and Shibagaki, S and Shimizu, R and Shimoda, T and Shimode, K and Shinkai, H and Shishido, T and Shoda, A and Somiya, K and Son, E J and Sotani, H and Sugimoto, R and Suzuki, T and Suzuki, T and Tagoshi, H and Takahashi, H and Takahashi, R and Takamori, A and Takano, S and Takeda, H and Takeda, M and Tanaka, H and Tanaka, K and Tanaka, K and Tanaka, T and Tanaka, T and Tanioka, S and Tapia San Martin, E N and Telada, S and Tomaru, T and Tomigami, Y and Tomura, T and Travasso, F and Trozzo, L and Tsang, T and Tsubono, K and Tsuchida, S and Tsuzuki, T and Tuyenbayev, D and Uchikata, N and Uchiyama, T and Ueda, A and Uehara, T and Ueno, K and Ueshima, G and Uraguchi, F and Ushiba, T and van Putten, M H P M and Vocca, H and Wang, J and Wu, C and Wu, H and Wu, S and Xu, W- R and Yamada, T and Yamamoto, K and Yamamoto, K and Yamamoto, T and Yokogawa, K and Yokoyama, J and Yokozawa, T and Yoshioka, T and Yuzurihara, H and Zeidler, S and Zhao, Y and Zhu, Z -H},
    title = "{Overview of KAGRA: Detector design and construction history}",
    journal = {Progress of Theoretical and Experimental Physics},
    volume = {2021},
    number = {5},
    pages = {05A101},
    year = {2020},
    month = {08},
    issn = {2050-3911},
    doi = {10.1093/ptep/ptaa125},
    url = {https://doi.org/10.1093/ptep/ptaa125},
    eprint = {https://academic.oup.com/ptep/article-pdf/2021/5/05A101/37974994/ptaa125.pdf},
}

@article{akutsu_kagra_2019,
	title = {{KAGRA}: 2.5 generation interferometric gravitational wave detector},
	volume = {3},
	issn = {2397-3366},
	url = {https://doi.org/10.1038/s41550-018-0658-y},
	doi = {10.1038/s41550-018-0658-y},
	number = {1},
	journal = {Nature Astronomy},
	author = {Akutsu, T. and Ando, M. and Arai, K. and Arai, Y. and Araki, S. and Araya, A. and Aritomi, N. and Asada, H. and Aso, Y. and Atsuta, S. and Awai, K. and Bae, S. and Baiotti, L. and Barton, M. A. and Cannon, K. and Capocasa, E. and Chen, C-S. and Chiu, T-W. and Cho, K. and Chu, Y-K. and Craig, K. and Creus, W. and Doi, K. and Eda, K. and Enomoto, Y. and Flaminio, R. and Fujii, Y. and Fujimoto, M.-K. and Fukunaga, M. and Fukushima, M. and Furuhata, T. and Haino, S. and Hasegawa, K. and Hashino, K. and Hayama, K. and Hirobayashi, S. and Hirose, E. and Hsieh, B. H. and Huang, C-Z. and Ikenoue, B. and Inoue, Y. and Ioka, K. and Itoh, Y. and Izumi, K. and Kaji, T. and Kajita, T. and Kakizaki, M. and Kamiizumi, M. and Kanbara, S. and Kanda, N. and Kanemura, S. and Kaneyama, M. and Kang, G. and Kasuya, J. and Kataoka, Y. and Kawai, N. and Kawamura, S. and Kawasaki, T. and Kim, C. and Kim, J. and Kim, J. C. and Kim, W. S. and Kim, Y.-M. and Kimura, N. and Kinugawa, T. and Kirii, S. and Kitaoka, Y. and Kitazawa, H. and Kojima, Y. and Kokeyama, K. and Komori, K. and Kong, A. K. H. and Kotake, K. and Kozu, R. and Kumar, R. and Kuo, H-S. and Kuroyanagi, S. and Lee, H. K. and Lee, H. M. and Lee, H. W. and Leonardi, M. and Lin, C-Y. and Lin, F-L. and Liu, G. C. and Liu, Y. and Majorana, E. and Mano, S. and Marchio, M. and Matsui, T. and Matsushima, F. and Michimura, Y. and Mio, N. and Miyakawa, O. and Miyamoto, A. and Miyamoto, T. and Miyo, K. and Miyoki, S. and Morii, W. and Morisaki, S. and Moriwaki, Y. and Morozumi, T. and Musha, M. and Nagano, K. and Nagano, S. and Nakamura, K. and Nakamura, T. and Nakano, H. and Nakano, M. and Nakao, K. and Narikawa, T. and Naticchioni, L. and Quynh, L. Nguyen and Ni, W.-T. and Nishizawa, A. and Obuchi, Y. and Ochi, T. and Oh, J. J. and Oh, S. H. and Ohashi, M. and Ohishi, N. and Ohkawa, M. and Okutomi, K. and Ono, K. and Oohara, K. and Ooi, C. P. and Pan, S-S. and Park, J. and Arellano, F. E. Peña and Pinto, I. and Sago, N. and Saijo, M. and Saitou, S. and Saito, Y. and Sakai, K. and Sakai, Y. and Sakai, Y. and Sasai, M. and Sasaki, M. and Sasaki, Y. and Sato, S. and Sato, N. and Sato, T. and Sekiguchi, Y. and Seto, N. and Shibata, M. and Shimoda, T. and Shinkai, H. and Shishido, T. and Shoda, A. and Somiya, K. and Son, E. J. and Suemasa, A. and Suzuki, T. and Suzuki, T. and Tagoshi, H. and Tahara, H. and Takahashi, H. and Takahashi, R. and Takamori, A. and Takeda, H. and Tanaka, H. and Tanaka, K. and Tanaka, T. and Tanioka, S. and Martin, E. N. Tapia San and Tatsumi, D. and Tomaru, T. and Tomura, T. and Travasso, F. and Tsubono, K. and Tsuchida, S. and Uchikata, N. and Uchiyama, T. and Uehara, T. and Ueki, S. and Ueno, K. and Uraguchi, F. and Ushiba, T. and van Putten, M. H. P. M. and Vocca, H. and Wada, S. and Wakamatsu, T. and Watanabe, Y. and Xu, W-R. and Yamada, T. and Yamamoto, A. and Yamamoto, K. and Yamamoto, K. and Yamamoto, S. and Yamamoto, T. and Yokogawa, K. and Yokoyama, J. and Yokozawa, T. and Yoon, T. H. and Yoshioka, T. and Yuzurihara, H. and Zeidler, S. and Zhu, Z.-H. and {KAGRA collaboration}},
	month = jan,
	year = {2019},
	pages = {35--40},
}

@article{copper1,
title = {Properties of copper and copper alloys at cryogenic temperatures. Final report},
author = {Simon, N. J. and Drexler, E. S. and Reed, R. P.},
doi = {10.2172/5340308},
url = {https://www.osti.gov/biblio/5340308},
journal = {National Institute of Standards and Technology},
place = {United States},
year = {1992},
month = {2}
}

@article{silicon-expansion,
    author = {Swenson, C. A.},
    title = "{Recommended Values for the Thermal Expansivity of Silicon from 0 to 1000 K}",
    journal = {Journal of Physical and Chemical Reference Data},
    volume = {12},
    number = {2},
    pages = {179-182},
    year = {1983},
    month = {04},
    issn = {0047-2689},
    doi = {10.1063/1.555681},
    url = {https://doi.org/10.1063/1.555681},
}

@Article{Hunter:2007,
  Author    = {Hunter, J. D.},
  Title     = {Matplotlib: A 2D graphics environment},
  Journal   = {Computing in Science \& Engineering},
  Volume    = {9},
  Number    = {3},
  Pages     = {90--95},
  publisher = {IEEE COMPUTER SOC},
  doi       = {10.1109/MCSE.2007.55},
  year      = 2007
}

@website{uncertainties,
    Title = {Uncertainties: a Python package for calculations with uncertainties},
    Author = {Lebigot, Eric O.},
    howpublished = "\url{http://pythonhosted.org/uncertainties/}",
}

@article{nemo,
    title = {Neutron Star Extreme Matter Observatory: A kilohertz-band gravitational-wave detector in the global network},
    volume = {37},
    DOI = {10.1017/pasa.2020.39},
    journal = {Publications of the Astronomical Society of Australia},
    publisher = {Cambridge University Press},
    author = {Ackley, K. and Adya, V. B. and Agrawal, P. and Altin, P. and Ashton, G. and Bailes, M. and Baltinas, E. and Barbuio, A. and Beniwal, D. and Blair, C. and et al.},
    year = {2020},
    pages = {e047}
}

@article{adhikari2020cryogenic,
	doi = {10.1088/1361-6382/ab9143},
	url = {https://doi.org/10.1088%2F1361-6382%2Fab9143},
	year = 2020,
	month = {jul},
	publisher = {{IOP} Publishing},
	volume = {37},
	number = {16},
	pages = {165003},
	author = {Adhikari, R.X. and Arai, K. and Brooks, A. F. and Wipf, C. and others},
	title = {A cryogenic silicon interferometer for gravitational-wave detection},
	journal = {Classical and Quantum Gravity},
}

@article{buikema2020sensitivity,
  title={Sensitivity and performance of the Advanced LIGO detectors in the third observing run},
  author={Buikema, A. and Cahillane, C. and Mansell, G. L. and Blair, C. D. and Abbott, R. and
  Adams, C.
  and Adhikari, R. X. and Ananyeva, A. and Appert, S. and Arai, K. and others},
  journal={Physical Review D},
  volume={102},
  number={6},
  pages={062003},
  year={2020},
  publisher={APS}
}

@inproceedings{acernese2020advanced,
  title={Advanced Virgo Status},
  author={Acernese, F. and Adams, T. and Agatsuma, K. and Aiello, L. and Allocca, A. and Amato, A.
 and Antier, S. and Arnaud, N. and Ascenzi, S. and Astone, P. and others},
  booktitle={Journal of Physics: Conference Series},
  volume={1342},
  number={1},
  pages={012010},
  year={2020},
  organization={IOP Publishing}
}

@article{abernathy2011einstein,
  title={Einstein gravitational wave Telescope conceptual design study},
  author={Abernathy, M. and Acernese, F. and Ajith, P. and Allen, B. and Amaro Seoane, P. and
  Andersson, N. and Aoudia, S. and Astone, P. and Krishnan, B. and Barack, L. and others},
  year={2011},
  publisher={EGO}
}

@whitepaper{ET2020einstein,
  title={Einstein Telescope Design Report Update 2020},
  author={ET Steering Committee Editorial Team},
  year={2020},
}

@article{reitze2019cosmic,
	author = {Reitze, D. and Adhikari, R. X. and Ballmer, S. and Barish, B. and
    Barsotti, L. and Billingsley, G. and Brown, D. A. and Chen, Y. and Coyne,
    D. and Eisenstein, R. and Evans, M. and Fritschel, P. and Hall, E. D. and
    Lazzarini, A. and Lovelace, G. and Read, J. and Sathyaprakash, B. S. and
    Shoemaker, D. and Smith, J. and Torrie, C. and Vitale, S. and Weiss, R. and
 Wipf, C. and Zucker, M.},
	journal = {Bulletin of the AAS},
	number = {7},
	year = {2019},
	month = {sep 30},
	note = {https://baas.aas.org/pub/2020n7i035},
	publisher = {},
	title = {Cosmic {Explorer}: The {U}.{S}. {Contribution} to {Gravitational}-{Wave} {Astronomy} beyond {LIGO}},
	volume = {51},
}

@whitepaper{team2020ETpathfinder,
  title={ETpathfinder Design Report},
  author={The ETpathfinder Team},
  year={2020},
}

@inbook{gombosi1994gaskinetic,
  title={Gaskinetic theory},
  author={Gombosi, Tamas I and Gombosi, Atmo},
  number={9},
  chapter={7.2.3},
  year={1994},
  publisher={Cambridge University Press}
}

@article{aasi2015advanced,
  title={Advanced ligo},
  author={Aasi, J. and Abbott, B. P. and Abbott, R. and Abbott, T. and Abernathy, M. R.
  and Ackley, K. and Adams, C. and Adams, T. and Addesso, P. and Adhikari, R. X. and
  others},
  journal={Classical and quantum gravity},
  volume={32},
  number={7},
  pages={074001},
  year={2015},
  publisher={IOP Publishing}
}

@article{acernese2014advanced,
  title={Advanced Virgo: a second-generation interferometric gravitational wave detector},
  author={Acernese, F. and Agathos, M. and Agatsuma, K. and Aisa, D. and Allemandou, N. and
  Allocca, A. and Amarni, J. and Astone, P. and Balestri, G. and Ballardin, G. and others},
  journal={Classical and Quantum Gravity},
  volume={32},
  number={2},
  pages={024001},
  year={2014},
  publisher={IOP Publishing}
}

@article{somiya2012detector,
  title={Detector configuration of KAGRA -- the Japanese cryogenic gra\-vitational-wave detector},
  author={Somiya, Kentaro},
  journal={Classical and Quantum Gravity},
  volume={29},
  number={12},
  pages={124007},
  year={2012},
  publisher={IOP Publishing}
}

@article{teflonconductivity,
    author = {Choi, Yeon and Kim, Dong Lak},
    year = {2012},
    month = {07},
    pages = {},
    title = {Thermal property of insulating material at cryogenic temperature},
    volume = {26},
    journal = {Journal of Mechanical Science and Technology},
    doi = {10.1007/s12206-012-0528-y}
}

@article{trott_experimental_2011,
    title = {An experimental assembly for precise measurement of thermal accommodation coefficients},
    volume = {82},
    issn = {0034-6748},
    url = {https://doi.org/10.1063/1.3571269},
    doi = {10.1063/1.3571269},
    pages = {035120},
    number = {3},
    journaltitle = {Review of Scientific Instruments},
    author = {Trott, Wayne M. and Castañeda, Jaime N. and Torczynski, John R. and Gallis, Michael A. and Rader, Daniel J.},
    date = {2011-03},
}

@article{ligovoyager1,
  title = {Prospects for doubling the range of Advanced LIGO},
  author = {Miller, John and Barsotti, Lisa and Vitale, Salvatore and Fritschel, Peter and Evans, Matthew and Sigg, Daniel},
  journal = {Phys. Rev. D},
  volume = {91},
  issue = {6},
  pages = {062005},
  numpages = {6},
  year = {2015},
  month = {Mar},
  publisher = {American Physical Society},
  doi = {10.1103/PhysRevD.91.062005},
  url = {https://link.aps.org/doi/10.1103/PhysRevD.91.062005}
}

@whitepaper{voyagerwhitepaper,
  title={LIGO Voyager Upgrade: Design Concept},
  author={Adhikari, Rana X. and Brooks, Aidan and Shapiro, Brett and McClelland, David and
  Gustafson, Eric
K. and Mitrofanov, Valery and Arai, Koji and Wipf, Christopher and Bonilla, Edgard},
  year={2023},
  url={https://docs.ligo.org/voyager/voyagerwhitepaper/main.pdf}
}

\end{document}